\documentclass[12pt]{article}
\usepackage{graphicx}
\usepackage{amsmath2000}

\def\hybrid{\topmargin 0pt      \oddsidemargin 0pt
        \headheight 0pt \headsep 0pt

       \textwidth 6.5in        
       \textheight 9in         
        \marginparwidth 0.0in
        \parskip 5pt plus 1pt   \jot = 1.5ex}
\catcode`\@=11
\def\marginnote#1{}

\newcount\hour
\newcount\minute
\newtoks\amorpm
\hour=\time\divide\hour by60
\minute=\time{\multiply\hour by60 \global\advance\minute by-\hour}
\edef\standardtime{{\ifnum\hour<12 \global\amorpm={am}%
        \else\global\amorpm={pm}\advance\hour by-12 \fi
        \ifnum\hour=0 \hour=12 \fi
        \number\hour:\ifnum\minute<10 0\fi\number\minute\the\amorpm}}
\edef\militarytime{\number\hour:\ifnum\minute<10 0\fi\number\minute}

\def\draftlabel#1{{\@bsphack\if@filesw {\let\thepage\relax
   \xdef\@gtempa{\write\@auxout{\string
      \newlabel{#1}{{\@currentlabel}{\thepage}}}}}\@gtempa
   \if@nobreak \ifvmode\nobreak\fi\fi\fi\@esphack}
        \gdef\@eqnlabel{#1}}
\def\@eqnlabel{}
\def\@vacuum{}

\def\draftmarginnote#1{\marginpar{\raggedright\scriptsize\tt#1}}

\def\draftlabel#1{{\@bsphack\if@filesw {\let\thepage\relax
   \xdef\@gtempa{\write\@auxout{\string
              \newlabel{#1}{{\@currentlabel}{\thepage}}}}}\@gtempa
   \if@nobreak \ifvmode\nobreak\fi\fi\fi\@esphack}
        \gdef\@eqnlabel{#1}}
\def\@eqnlabel{}
\def\@vacuum{}
\def\draftmarginnote#1{\marginpar{\raggedright\scriptsize\tt#1}}

\def\draft{\oddsidemargin -.5truein
        \def\@oddfoot{\sl preliminary draft \hfil
        \rm\thepage\hfil\sl\today\quad\militarytime}
        \let\@evenfoot\@oddfoot \overfullrule 3pt
        \let\label=\draftlabel
        \let\marginnote=\draftmarginnote
   \def\@eqnnum{(\theequation)\rlap{\kern\marginparsep\tt\@eqnlabel}%
\global\let\@eqnlabel\@vacuum}  }

\def\numberbysection{\@addtoreset{equation}{section}
        \def\theequation{\thesection.\arabic{equation}}}

\def\underline#1{\relax\ifmmode\@@underline#1\else
        $\@@underline{\hbox{#1}}$\relax\fi}

\def\titlepage{\@restonecolfalse\if@twocolumn\@restonecoltrue\onecolumn
     \else \newpage \fi \thispagestyle{empty}\c@page\z@
        \def\thefootnote{\fnsymbol{footnote}} }

\def\endtitlepage{\if@restonecol\twocolumn \else  \fi
        \def\thefootnote{\arabic{footnote}}
        \setcounter{footnote}{0}}  
\relax

\draft

\numberbysection
\hybrid

\def\beq{\begin{equation}}
\def\eeq{\end{equation}}
\def\[{\left[}
\def\]{\right]}
\def\({\left(}
\def\){\right)}
\def\pa{\partial}
\def\m{\mu}
\def\d{\delta}
\def\g{\gamma}
\def\l{\lambda}
\def\s{\sigma}
\def\t{\theta}
\def\e{\epsilon}
\def\p{\pi}
\def\Ker{\rm Ker}
\def\C{\mathbf{C}}
\def\Rn{Rat_{N}}
\def\spec{\rm{spec}}
\def\SL{{\cal S}(L_0)} 
\def\bea{\begin{eqnarray}}
\def\eea{\end{eqnarray}}
\newtheorem{thm}{Theorem}

\begin{document}

\begin{titlepage}

\title{The Atiyah--Hitchin bracket and the open Toda lattice}

\author{K.L. Vaninsky \thanks{Department of Mathematics, 
Michigan State Unversity, East Lansing MI 48824 
This work  is partially supported  by NSF grant
DMS-9971834.  The author would like to thank MPI Bonn
where the paper was written  for hospitality.} }
\date{July 17, 2001}
\maketitle

\begin{abstract} 
The dynamics of finite nonperiodic Toda lattice is an isospectral 
deformation of the finite three--diagonal Jacobi matrix. It is known 
since the work of Stieltjes that such matrices are in one--to--one 
correspondence with their Weyl functions. These  
are rational functions mapping the upper half--plane into itself. 
We consider   representations of the Weyl functions 
as a quotient of two polynomials and  
exponential representation.   We establish  a connection  between 
these representations and recently developed algebraic--geometrical approach 
to the inverse problem for Jacobi matrix.  
The space of rational functions has natural 
Poisson structure discovered by Atiyah and Hitchin. 
We show that an invariance of the AH structure  
under linear--fractional transformations leads to  two systems 
of canonical coordinates and two families of commuting Hamiltonians.
We establish  a relation  of one of these systems with Jacobi elliptic 
coordinates. 
\end{abstract}

\end{titlepage}
\newpage

\section{Introduction.} 
The Toda lattice is a mechanical system of $N$--particles connected by  
elastic  strings.
The Hamiltonian of the system is 
$$
H=\sum\limits_{k=0}^{N-1}\frac{p_k^2}{ 2} +\sum\limits_{k=0}^{N-2}e^{q_k-q_{k+1}}.
$$
Introducing the classical Poisson bracket
$$
\{f,g\}=\sum\limits_{k=0}^{N-1} \frac{\pa f}{\pa q_k} \frac{\pa g}{ \pa p_k}- 
                                \frac{\pa f}{ \pa p_k} \frac{\pa g}{\pa q_k},  
$$
we write the equations of motion as  
\begin{eqnarray}
q_k^{\bullet}&=&\{q_k,H\}= p_k, \nonumber\\
p_k^{\bullet}&=&\{p_k,H\}=-e^{q_k-q_{k+1}}+e^{q_{k-1}-q_{k}},\qquad\qquad\qquad
k=1,\ldots,N-1. \nonumber
\end{eqnarray}
We put $q_{-1}=-\infty,\;  q_N=\infty$ in all formulae. 
Following \cite{F,M} introduce the new variables
$$
c_k=e^{q_k-q_{k+1}/2},\quad\qquad\qquad v_k=-p_k.
$$
In these variables
$$
H=\sum\limits_{k=0}^{N-1}\frac{v_k^2}{ 2} +\sum\limits_{k=0}^{N-2}c_k^2,
$$
and 
\beq\label{lps}
\{c_k,v_k\}=-c_k/ 2,\qquad\qquad\{c_k,v_{k+1}\}=c_k/ 2. 
\eeq
The equations of motion take the form
\begin{eqnarray}
v_k^{\bullet}&=& \{v_k,H\}=c_k^2-c_{k-1}^2,\nonumber\\
c_k^{\bullet}&=& \{c_k,H\}=c_k(v_{k+1}-v_{k})/2 .\nonumber
\end{eqnarray}
These equations are compatibility conditions  for the Lax equation $L^{\bullet}=\[A,L\]$,  where 
$$
L=\left[\begin{array}{ccccc}
v_0      & c_0       & 0         &\cdots        & 0\\
c_0      & v_1       & c_1       &\cdots        & 0\\
\cdot        &  \cdot    & \cdot     &\cdot         & \cdot\\
0    &  \cdots         & c_{N-3}         & v_{N-2}     & c_{N-2}\\
0       & \cdots    & 0         & c_{N-2}      & v_{N-1}
\end{array}\right]
$$
and 
$$
2A=\left[\begin{array}{ccccc}
0        & c_0       & 0         &\cdots        & 0\\
-c_0     & 0         & c_1       &\cdots        & 0\\
\cdot        &  \cdot    & \cdot     &\cdot         & \cdot\\
0    &  \cdots         & - c_{N-3}         & 0      & c_{N-2}\\
0       & \cdots    & 0         & - c_{N-2}      &       0 
\end{array}\right]. 
$$
The Lax formula implies that the spectrum $\l_0<\ldots < \l_{N-1}$ 
of $L$ remain fixed. 
It is known since the work of Stiltjies, \cite{ST}, that the rational  function 
$w(\l)=(R(\l)\d(0),\d(0))$ where $R(\l)=(L-\l I)^{-1}$ is the resolvent,   
plays the key role in reconstruction of the 
matrix $L$ from its spectral data.  It was encountered  later in the spectral 
theory of the Sturm-Liouwille operator, \cite{W}, and received the name 
of  Weyl function.  Simply by expanding it in the  continued fraction
\begin{equation}\nonumber
w(\l)=- \cfrac{1}{\l-v_0-
         \cfrac{c_0^2}{\l-v_1-
          \cfrac{c_1^2}{\l-v_2- \dotsb {
           \cfrac{c_{N-2}^2}{\l-v_{N-1}
}}}}}
\end{equation}
one can read the entries of $L$. This fact is central to  Moser's  solution 
of the nonperiodic Toda, \cite{MO}. At the same time Stiltjies method appears 
to be  a computational trick and does not provide a conceptual explanation 
why the function $w(\l)$ actually determines $L$. 

An attempt to  solve an inverse spectral problem for the finite Jacobi matrix 
using algebraic--geometrical approach was made at the beginning of  eighties by 
McKean, \cite{MA}.   
It was realized  that 
the corresponding  spectral curve  is a singular reducible Riemann surface. 

The  recent  interest in the spectral curves of finite   
Toda lattice  stems from various sources. One is  the 
work of Seiberg and Witten, \cite{SW1,SW2} on sypersymmetric Yang--Mills 
theories. The smooth hyperelliptic spectral curves of the periodic Toda 
appear in the pure gauge $N=2$ SUSY Yang--Mills models in $4$ dimensions 
\cite{GK}.  The  physically interesting limit of the theory which corresponds 
to transition from  the smooth hyperelliptic curve  to the singular one  
of \cite{MA} was considered  in  \cite{BM}. 
Another source of interest is  the  Camassa--Holm equation, \cite{CH}.   
Is is known, see \cite{BR, BSS},  that the dynamics of, so--called 
peakons solutions, is isomorphic   to an  isospectral flow on the space of 
finite Jacobi matrices.  

A solution of nonperiodic Toda within  algebraic--geometrical  approach was 
obtained recently in \cite{KV}.  The poles of $w(\l)$ determine the curve 
and the zeros specify the divisor   of the Baker--Akhiezer function. 
This information uniqly specifies  the 
BA function and therefore the matrix $L$. Whence that fact that $w(\l)$ 
determines the matrix can be considered as a consequence of the Riemann--Roch 
theorem which guarantees uniqness of the BA function.
Using   BA functions    we obtained  in \cite{KV} the  
explicit formulae for the solution, the symplectic structure 
and two systems of  canonical   coordinates for it. 

The present paper combines  ideas developed in \cite{V} with   
algebraic--geometrical approach of \cite{KV}.  
It can be divided in two parts. In the first part of the paper (sections 2--5) 
we establish relations between  the standard 
objects of spectral theory and    algebraic-geometrical constructions.
We show how the BA function can be constructed from the 
orthogonal polynomial of the first kind and suitably normalized Weyl solution. 
The  rational functions which map the upper half--plane
into itself are  defined  by the formula  
\beq\label{hweil}
w(\l)= \frac{\rho_0}{ \l_0-\l}+\ldots +\frac{\rho_{N-1}}{ \l_{N-1}-\l}, \qquad
\qquad \rho_k > 0. 
\eeq
 They are parametrized by $2N$ parameters $\rho$'s and $\l$'s. 
The rational functions
corresponding to the Weyl functions of finite Jacobi matrices are
specified  by the condition $\sum \rho_k =1$.

We consider different reperentations of the Weyl functions. 
The first one, the  ratio of two monic polynomials, 
$$
w(\l)=-\frac{q(\l)}{ p(\l)}
$$
leads to the standard Abel map. Another, the exponential representation, 
which employs  Krein's  spectral shift function $\xi(z)$ 
$$
w(\l)=\frac{1}{\l_0-\l} \exp \Xi(\l),  \qquad\qquad
 \Xi(\l)=\int \frac{\xi(z)}{ z-\l} dz
$$
produces the Abel map in Baker's form.

In the second part (sections 6--9) we study  the Atiyah---Hitchin bracket 
on rational functions and its relation to the Hamiltonian formalism for the 
Toda lattice. 
As it was  recently discovered by Faybusovich and 
Gehtman, \cite{FG}  the AH bracket on rational  functions can be written as 
\beq\label{lpsah}
\{w(\l),w(\m)\}=\frac{(w(\l)-w(\mu))^2}{ \l-\m}
\eeq
The bracket \ref{lps}  corresponds 
to the restriction of \ref{lpsah} on the   $2N-1$ dimensional submanifold 
corresponding to Weyl functions of $N\times N$ Jacobi matrices. 
We show  that the restricted bracket has 
two systems of canonical coordinates on symplectic leafs of the foliation 
defined by  level sets of the Casimir $\sum\l_k$. The  existence of 
these  two canonical coordinate  systems  is a consequence of  
invariance of the AH bracket  under the group 
of linear--fractional transformations
$$
w\quad  \longrightarrow \quad w'=\frac{aw+b}{ cw+d}.
$$

The first system of canonical coordinates is 
associated with $N$  poles of $w(\l):\quad \l_0<\ldots < \l_{N-1}$. 
The half of variables is $N-1$ points of the spectrum $\l_1,\hdots,\l_{N-1}$. 
Another half are the functions  of $q(\l)$   at these points
$$
\theta_k=\log \frac{(-1)^k q(\l_k)}{ q(\l_0)},\qquad\qquad k=1,\hdots,N-1.
$$ 
This system is called action--angles coordinates. The  
verification of canonical relations  is  obtained by computing residues. 
The associated Hamiltonias are 
$$
H_j=\frac{1}{ j}\sum \l_n^j,\qquad \qquad\qquad j=1,\ldots,N. 
$$ 
They produce the standard Toda flows which  preserve the spectrum.  

The second system  is coming from $N-1$  finite roots $\g_1<\ldots <\g_{N-1}$ 
of the equation  
$$
w(\g)= \frac{\rho_0}{ \l_0-\g}+\ldots +\frac{\rho_{N-1}}{ \l_{N-1}-\g}=0.
$$
The adjoint $N-1$ variables are 
$$
\pi_k=\log (-1)^{N+k} p(\gamma_k),\qquad\qquad k=1,\hdots,N-1. 
$$
These are  the divisor--quasimomentum coordinates. The Hamiltonians of this
system are 
$$
T_j=\frac{1}{ j}\sum \g_n^j,\qquad\qquad\qquad j=1,\ldots,N-1.
$$
These Hamiltonians produce the flows transversal to the isospectral 
manifolds. These flows preserve the divisor.

Now we will explain the origins of the divisor--quasimomentum 
coordinate system. 
Jacobi, \cite{Jac} lecture 26, introduced his famous elliptic coordinates 
as  finite roots $\g_0< \g_1<\ldots <\g_{N-1}$   of the equation
$$
w(\g)= \frac{\rho_0}{ \l_0-\g}+\ldots +\frac{\rho_{N-1}}{ \l_{N-1}-\g}=1.
$$ 
These $N$ roots  are considered to be the functions of $N$ independent 
$\rho$'s while $\l$'s  remain fixed. 
In the case of finite Jacobi matrices this choice of  parameters $\g$ 
is bad. Due to the constraint $\sum\rho_k=1$ they are functionaly dependent. 
The same is true for any other constant instead of $1$. Only for the special 
value  $0$ the equation $w(\g)=0$ has one "unmovable" root $\g_0=\infty$ and 
all other $N-1$ finite roots are functionaly independent.  Whence 
the divisor can be considered as a variant  of the Jacobi elliptic coordinates. 

The transformation $ w(\l)\rightarrow w'(\l)=-1/w(\l)$ defines the dual Weyl 
function $w'$. The roots of the equations $w(\g)=0$ become poles of the 
function $w'(\l)$.  
The invariance of the AH structure allows to establish canonical 
character of the divisor--quasimomentum coordinates  again by computing 
residues.

We would like to conclude the introduction with the following remark.  
The tradional study of the integrable dynamics is a part of Hamiltonian 
mechanics with its' standard objects like Poisson and symplectic manifolds, 
vector fields, differential forms,   {\it etc}. 
We  demonstrate  in this paper that  the integrable dynamics on the space of  
Jacobi matrices or equivalently Weyl functions can be 
reformulated and studied  purely in terms of complex analysis.

{\it Organization of the paper}. In  section 2 we review  standard 
constructions of spectral theory of three--diagonal Jacobi matrices. 
In section 3 
we introduce the spectral shift function and establish the trace formulae. 
In section 4 we consider the reducible Riemann surface and show 
how the Baker-Akhiezer function can be constructed using the orthogonal 
polynomial of the first kind and suitably normalized Weyl solution.
In section 5 we consider representation of the Weyl function as a 
ratio of two polynomials and exponential representation. We demonstrate how 
using these representations one can construct the Abel map in the 
ordinary form and in the Baker form, correspondingly. We also describe  the 
range of the map and show its' one--to--one character. This section 
completes the description of the spectral theory and its' relation to 
algeraic geometry. In section 6 we introduce the Atiyah--Hitchin  Poisson 
bracket 
and  describe  some of its' elementary properties. In the next,  7-th
section we construct canonical  variables for the AH bracket. 
We also construct the 
Dirac restriction of the AH bracket on the submanifod of  Weyl functions.  
We introduce  two systems of canonical coordinates for the restricted bracket  
in section 8. We also linearize the flow  in terms of spectral shift function. 
The last 9 section describes the isospectral and transversal flows. 

{\it Acknoledgements.} The author  thank H. Braden, H. McKean, S. Natanzon 
and I. Kriche\-ver for stimulating discussions.

\section{The Direct Spectral Problem.}
Most of the material of this section can be found in \cite{A3} and presented 
here in order to set notations. Consider a  finite Jacobi matrix
$$ 
L=\left[\begin{array}{ccccc}
v_0      & c_0       & 0         &\cdots        & 0\\
c_0      & v_1       & c_1       &\cdots        & 0\\
\cdot        &  \cdot    & \cdot     &\cdot         & \cdot\\
0    &  \cdots         & c_{N-3}         & v_{N-2}     & c_{N-2}\\
0       & \cdots    & 0         & c_{N-2}      & v_{N-1}
\end{array}\right]. 
$$
It acts as a  selfadjoint operator  in the complex $l^2[0,N-1]$ with standard orthonormal basis 
$\d(k)=\underbrace{(\ldots,1,\ldots)}_{k\mbox{-th place}},\;  k=0,\ldots,N-1$. 
The operator $L$ has simple spectrum  
$\l_0 <    \ldots <  \l_{N-1}$ corresponding to normalized eigenvectors $e(\l_k)= 
(e_0(\l_k),\ldots, e_{N-1}(\l_k)),\; k=0,\ldots,N-1$. Let 
$E(\l)=\sum\limits_{\l_k <\l} e(\l_k)\otimes e(\l_k)$ be an orthogonal spectral measure of $L$.   
Thus 
\beq\label{op}
L= \int \l dE(z)
\eeq
and
\beq\label{res}
R(\l)= (L-\l I)^{-1} =\int\frac{dE(z)}{ z-\l}.
\eeq
For any two vectors $u,\;v$ the Parseval identity holds 
\beq\label{par}
(u,v)=\sum\limits_{k}(u,e(\l_k))(v,e(\l_k))
\eeq

We associate with $L$ the eigenvalue problem 
\begin{eqnarray}
v_0y_0 +c_0y_1&=& \l y_0, \label{zer}\\
c_{0}y_{0}+v_1y_1 +c_1y_2&=& \l y_1, \label{fe}\\ 
c_{n-1}y_{n-1}+v_ny_n +c_ny_{n+1}&=& \l y_n,\qquad n =2,\ldots,N-2;\nonumber \\
c_{N-2}y_{N-2}+v_{N-1}y_{N-1} +c_{N-1}y_{N}&=& \l y_{N-1}.\label{la}
\end{eqnarray}
The coefficient $c_{N-1}$ is defined by the formula $c_{N-1} =\prod_{k=0}^{N-2}  c_k^{-1} $. 
For the system \ref{zer}-\ref{la} we introduce the solution
$$
P(\l):\qquad\qquad P_{-1}(\l)=0,\; P_0(\l)=1, \;\ldots,P_N(\l);
$$
and for  the system \ref{fe}-\ref{la}
$$
Q(\l):\qquad\qquad  Q_0(\l)=0,\; Q_1(\l)=\frac{1}{ c_0},\ldots, Q_N(\l).
$$
Let $L_{[k,p]}$ be the truncated matrix
\beq\label{mat}
L_{[k,p]}=\left[\begin{array}{ccccc}
v_k      & c_k          &\cdots        & 0\\
c_k      & v_{k+1}       &\cdots        & 0\\
\cdots   & \cdots       & \ddots    &c_{p-1}\\
\cdots   & \cdots       & c_{p-1}      & v_{p}
\end{array}\right]. 
\eeq
Then for $n=1,2,\ldots,N-1$ 
\beq
P_n(\l)=(-1)^n\frac {\det(L_{[0,n-1]}-\l I)}{ \prod\limits_{k=0}^{n-1}c_k}
\label{pi}, 
\eeq
and 
\beq 
Q_n(\l)=(-1)^{n+1}\frac{\det(L_{[1,n-1]}-\l I )}{ \prod\limits_{k=0}^{n-1}c_k}.
\label{qu} 
\eeq 
Now we will use the solutions $P$ and $Q$ to give the formula \ref{par} more 
concrete form.

For $  P(\l)=(P_0(\l),\ldots,P_{N-1}(\l))$ from \ref{pi} we have 
\beq\label{kl}
e(\l_k)=P(\l_k)\, \sqrt{\rho_k}, \qquad\qquad\qquad \rho_k=\frac{1 }{
\sum\limits_{n=0}^{N-1} P^2_n(\l_k)}.
\eeq
Thus  \ref{par} takes the form
\beq\label{pars}
(u,v)=\int \tilde{u}(\l)\tilde{v}(\l)\,d\s(\l)
\eeq
where 
$$
\tilde u(\l)=(u,P(\l)),\qquad \tilde v(\l)=(v,P(\l)),\qquad  d\sigma(\l)=\sum\d(\l-\l_k)\rho_k.
$$ 
Moreover, using  \ref{kl} we have 
\beq\label{pith}
(e(\l_k), \d(0))=\sqrt{\rho_k} >0,\qquad\qquad\qquad \sum\limits_{k=0}^{N-1}\rho_k=1.
\eeq
This implies, in particular,  that  $\d(0)$ is a cyclic vector for $L$.

For the Weyl function defined as $$w(\l)=-\frac{Q_N(\l)}{ P_N(\l)}$$ formulae \ref{pi}-\ref{qu} imply
\beq\label{we}
w(\l)=-\frac{(-1)^{N+1}\det (L_{[1,N-1]}-\l I)}{ (-1)^N \det (L-\l I)}= 
-\frac{(-1)^{N+1}\prod\limits_{s=1}^{N-1} (\gamma_s-\l)}{(-1)^N\prod\limits_{n=0}^{N-1} (\l_n-\l)}, 
\eeq
where  the roots $\l$ and $\g$ interlace  
\beq\label{int}
\l_0 <  \g_1 <  \l_1 <   \ldots <   \l_{N-2} <  \g_{N-1} <  \l_{N-1}
\eeq
due to the Sturm theorem.

By construction $Q_N+ wP_N=0$ for all $\l$. Formulae  \ref{zer}-\ref{la} produce  
$(L -\l I) (Q+wP)=\d(0)$ and $Q+wP =R(\l) \d(0)$. The formula  \ref{res} implies 
\beq\label{weil}
w(\l)= (R(\l)\d(0), \d(0)) = \int\frac{(\d(0),dE (z)\d(0)) }{ z-\l}= 
\int\frac{d\s (z)}{ z-\l}. 
\eeq
From \ref{op} for the moments of the measure $d\s$ we have 
\beq\label{mom}
s_k=(L^k \d(0),\d(0))=\int \l^k\, d \s(\l).
\eeq
Using \ref{pith},  
\beq\label{as}
w(\l)=-   \sum\limits_{n=0}^{\infty} s_n \l^{-(n+1)},   \qquad\qquad\mbox{where}\qquad s_0=1. 
\eeq 

We conclude this section with derivation of trace formulae. 
To simplify notations  we assume that $\l_0=0$. Then, \ref{we} becomes
\beq\label{pr}
w(\l)=-\frac{1}{ \l}\prod_{k=1}^{N-1}\(\frac{\g_k-\l}{ \l_k-\l}\).
\eeq
After simple algebra, 
$$
\frac{\g_k-\l}{ \l_k-\l}= 1+\sum\limits_{p=1}^{\infty}\frac{\l^{p-1}_k(\l_k-\g_k)} 
{ \l^p} = \sum\limits_{p=0}^{\infty}\Delta_k^p \l^{-p},
\qquad \mbox{where}\qquad \Delta_k^0=1. 
$$
Fathermore,
$$
w(\l)=-\sum\limits_{n=0}^{\infty}
\[\sum\limits_{p_1+ \cdots+p_{N-1}=n}\prod_{k=1}^{N-1} \Delta_k^{p_k}\] \l^{-(n+1)}.
$$
Comparing  it with \ref{as} we obtain the trace formulae
\beq\label{Hoh}
s_n= \sum\limits_{p_1+ \cdots+p_{N-1}=n}\prod_{k=1}^{N-1} \Delta_k^{p_k}.
\eeq
The first few  are listed below
\bea
s_1&=& \sum\limits_k \Delta_k^1, \nonumber\\
s_2&=& \sum\limits_k \Delta_k^2 + \sum\limits_{k_1\neq k_2} \Delta_{k_1}^1 \Delta_{k_2}^1\nonumber,\\
s_3&=& \sum\limits_k \Delta_k^3 + 2 \sum\limits_{k_1\neq k_2} \Delta_{k_1}^2 \Delta_{k_2}^1
+\sum\limits_{k_1\neq k_2 \neq k_3} \Delta_{k_1}^1 \Delta_{k_2}^1 \Delta_{k_3}^1,\qquad etc. 
\nonumber
\eea

To derive  the standard  trace formulae, \cite{Ho},  
from  the resolvent   expansion
$$
R(\l)=(L-\l I)^{-1}=  -\frac{I}{ \l}-\frac{L }{ \l^2}-\frac{L^2}{ \l^3}-\ldots;
$$
we obtain 
\begin{eqnarray}\label{re}
w(\l)&=& -\frac{(I \d(0),  \d(0)) }{ \l} - \frac{(L \d(0), \d(0)) }{ \l^2} -
\frac{(L^2 \d(0), \d(0)) }{ \l^3}-\ldots \nonumber \\
&=&  -\frac{1}{ \l} - \frac{v_0}{ \l^2} - \frac{v_0^2 + c_0^2 }{ \l^3}-\ldots.
\end{eqnarray}
Matching the coefficients in \ref{as} and \ref{re},

\bea
v_0&=& \sum\limits_k \Delta_k^1,\nonumber\\
c_0^2&=& \sum\limits_k \Delta_k^2 - \sum\limits_{k} \(\Delta_{k}^1\)^2,\qquad etc. 
\nonumber
\eea

\section{The trace formulae via Krein spectral shift.} 
We assume $\l_0=0$, then \ref{we} becomes
\bea\label{rld}
w(\l)=-\frac{1}{ \l}\frac{\det (L_{[1,N-1]}-\l I)}{ \det (L\, |\,\mbox{Ker}\, 
L^\bot -\l I)}, 
\eea
where $L\, |\,\mbox{Ker}\, L^\bot$ ia a restriction of $L$ on the orthogonal compliment to $\mbox{Ker}\, L$. 
By elementary transformations the ratio of two determinants can be put in the form
\begin{eqnarray}\label{kss}
w(\l)& = &  -\frac{1}{ \l} \prod\limits_{s=1}^{N-1} \(\frac{\g_s-\l}{ \l_s-\l}\) 
 = -\frac{1}{ \l} \exp\sum\limits_{s=1}^{N-1} \int_{\l_s}^{\g_s} 
\frac{dz}{ z-\l}\nonumber \\ 
&=& -\frac{1}{ \l} \exp \Xi(\l) =  -\frac{1}{ \l} \exp {\int \frac{\xi(z)}{ z-\l} dz}, 
\end{eqnarray}
where\footnote{$n_{L}(z)$ is a counting 
function $n_{L}(z)=\#\{\mbox{eigenvalues of}\; L \leq z\}  $.} 
$$\xi(z)=n_{L\, |\,\Ker\, L^\bot}(z)- n_{L_{[1,N-1]}}(z)$$ is the 
Krein spectral shift 
function, \cite{K}. This  exponential 
representation of the 
Weyl function has much wider range of applicability then the formula \ref{rld}, 
which requires 
separate existence of determinants in the numerator and denominator.  It can be obtained,  for 
example,  for infinite unbounded matrices   
under very mild condition on  closeness of $L_{[1,\infty]}$ and  $L\,|\,\mbox{Ker}\, L^\bot$. 

One can obtain trace formulae in terms of $f_k=\int z^k \xi(z) dz$ entering into the asymptotic 
expansion
$$
\Xi(\l)= - \sum\limits_{n=0}^{\infty} f_n \l^{-(n+1)}.
$$ 
Expanding the exponent in \ref{kss} and matching  the coefficients 
with \ref{as} 
\bea
s_1&=& -f_0\nonumber,\\
s_2&=& \frac{f_0^2}{ 2}- f_1\nonumber,\\
s_3&=&  f_0f_1- f_2 -\frac{f_0^3}{ 6}, \qquad etc.
\nonumber
\eea
Evidently, these formulae can be put in the form \ref{Hoh} 
using representation of 
$\xi(z)$ as a difference of two counting functions.

\section{The Spectral Curve. The Baker-Akhiezer Function.}
The function $w(\l)$  determines the matrix $L$ or in another words 
the functions $w(\l)=w(\l, \,L)$ are coordinates on the space $\mathcal L$ of all $N\times N$ Jacobi
matrices. The "index" $\l$ which labels  the "coordinates" takes the values in 
$\C^1\,  \backslash \, \spec \;  L$. This statement goes back to Stieltjes, 
\cite{ST}. 
 There are  two standard ways to recover $L$ from $w$. The first is
to expand $w(\l)$ into continuous fraction, from which one can read the coefficients of $L$.
The second is to construct  polynomials orthogonal with respect to the  spectral
measure recovered from
$w(\l)$. A three term recurrent relation for these polynomials is, in fact, the matrix $L$.
Recently, the classical  inversion problem, received  a new,  algebro-geometrical 
solution, \cite{KV}. The main novel part of \cite{KV} is,   so--called, Baker-Akhiezer function  for a 
{\it reducible} curve. This construction is described below.

We start with the standard Weyl solution $Q+wP$ and note that $\frac{Q+wP}{ w}$ is a 
solution of \ref{fe}-\ref{la} which is equal to $1$ at $n=0$ and 
vanishes at $n=N$ for  all $\l$. The vector $P$ is also a solution of \ref{fe}-\ref{la} 
which is equal to $1$ for $n=0$ and 
all values of $\l$ and vanishes at $n=N$ for $\l=\l_k$.    Thus we have a "gluing" condition:
\beq\label{glue}
P(\l_k)=\frac{Q+wP}{ w}(\l_k).
\eeq  
In other words, at the points of the spectrum the function $w(\l)$ conjugates two solutions $P$ and 
$Q+wP$ which vanish at the left ($n=-1$) or right ($n=N$) correspondingly.

The  singular algebraic curve $\Gamma$ (Figure 1)   is obtained by gluing at 
the points of the 
spectrum two copies of the complex plane. 
\begin{figure}
\centering
\includegraphics[width=0.60\textwidth]{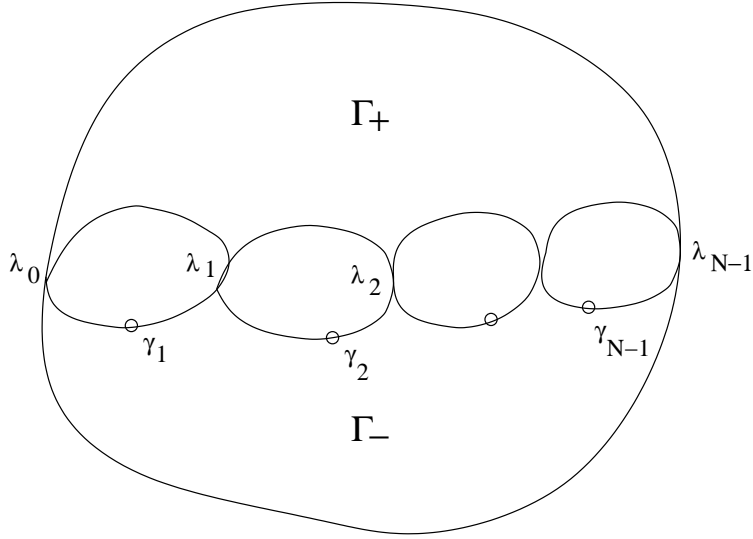}
\caption{Riemann surface}
\end{figure}
Define Baker-Akhiezer function on $\Gamma$ by the formula, $\l=\l(\e)$:
$$
\psi(\e)= \left\{
\begin{array}
  {r@{\quad:\quad}l}
P(\l)  &       \qquad\qquad  \mbox{if}\;\;     \e \in \Gamma_+;\\
Q+wP/ w(\l) &  \qquad\qquad  \mbox{if}\;\;      \e \in \Gamma_-.
\end{array} \right.  
$$
The function $\psi$ is continuous on $\Gamma$ due to the gluing condition \ref{glue}. The BA function  
has the only simple poles at the points of the divisor $(\gamma_1,-),\ldots,(\gamma_{N-1},-)$. At two 
infinities the BA function  has poles of prescribed order. These 
data  determine the BA function and therefore the operator $L$. Thus,  the uniqueness statement 
can be viewed as a consequence of the Riemann-Roch theorem. This remark 
completes the  description of the  direct spectral problem. The inverse 
problem can be solved using an  explicit formula for the 
time--dependent BA function (for details see
\cite{KV}).

\section{The Abel map and the Jacobian.}
Let  $Rat_N$ be a  set of all rational functions  wich map the upper half--plane into itself, vanish at 
infinity and have  $N$  poles. 
Any  function $w(\l)$ from $Rat_N$ has the form 
\beq\label{repr}
w(\l)=\sum\limits_{k=0}^{N-1} \frac{\rho_k }{  \l_k- \l}
\eeq
with real poles at $\l_0< \ldots < \l_{N-1}$ and $\rho_k>0$.  For $\l$ real below/above spectrum the
function $w(\l)$ is positive/negative. Furthermore,
$$
w'(\l)=\sum \frac{\rho_k}{ (\l_k-\l)^2} >0.
$$
and $w(\l)$ continuously changes from minus infinity to plus infinity, when $\l$ runs
between two consecutive poles. Thus the function $w(\l)$ has exactly $N-1$ zeros $\gamma$'s which
interlace $\l$'s as in the formula \ref{int}.

Furthermore,  any function  from $Rat_N$  can be represented  as a ratio of two polynomials
\beq\label{det}
w(\l)=-\frac{q(\l)}{ p(\l)} =-\frac{q_0(-1)^{N+1}\prod\limits_{s=1}^{N-1} 
(\gamma_s-\l)}{ (-1)^{N}\prod\limits_{n=0}^{N-1} (\lambda_n -\l)} =
-\frac{q_0 \l^{N-1} +  q_1 \l^{N-2} + \ldots + q_{N-1}}{
\l^{N} +  p_0 \l^{N-1} + \ldots + p_{N-1}}.
\eeq
The polynomials are  defined up to a multiple factor. In the formula \ref{det} it is chosen such that 
the leading coefficient of the denominator is $1$, similar to \ref{we}. Evidently, the polynomial 
$q(\l)$ can be determined from its' values $q(\l_0), \ldots, q(\l_{N-1})$ which are 
{\it free}  parameters. 

Now we turn to the submanifold $Rat_N'$ with $\sum\rho_k=1$.
These are the functions from $Rat_N$  which are Weyl functions of finite Jacobi matrices.
The values $q(\l_0),\ldots,q(\l_{N-1})$ are not independent anymore.
Indeed, from the identity
$$
-\frac{q(\l)}{ p(\l)}=\sum\frac{\rho_n}{ \l_n-\l}
$$
and condition $p(\l_n)=0$ we obtain
$
q(\l_k)=p'(\l_k)\rho_k.
$
Therefore, 
$$
\sum\limits_{n=0}^{N-1}\frac{q(\l_n)}{ p'(\l_n)}=1.
$$
Due to the relation $q_0=\sum \rho_k$, an  another way to say  that 
$w(\l) \in Rat_N'$ is  that the polynomial $q(\l)$ in  
\ref{det} is a monic polynomial.

For a function $w(\l) \in Rat'_N$ we define  angle variables by the formula
\beq\label{angl}
\theta_k=\log \frac{(-1)^k q(\l_k)}{ q(\l_0)}\qquad\qquad\qquad k=1,\ldots, N-1.
\eeq
The are exactly $k$ roots of $q(\l)$ between $\l_0$ and
$\l_k$ and  it changes sign $k$ times when $\l$ varies from $\l_0$ to $\l_k$. Whence
variables $\theta$'s are always real.

To clarify geometrical meaning of the variables $\t$'s we introduce (Figure 2) 
a standard homology basis 
on the curve $\Gamma$ corresponding to some Jacobi matrix.  
\begin{figure}
\centering
\includegraphics[width=0.60\textwidth]{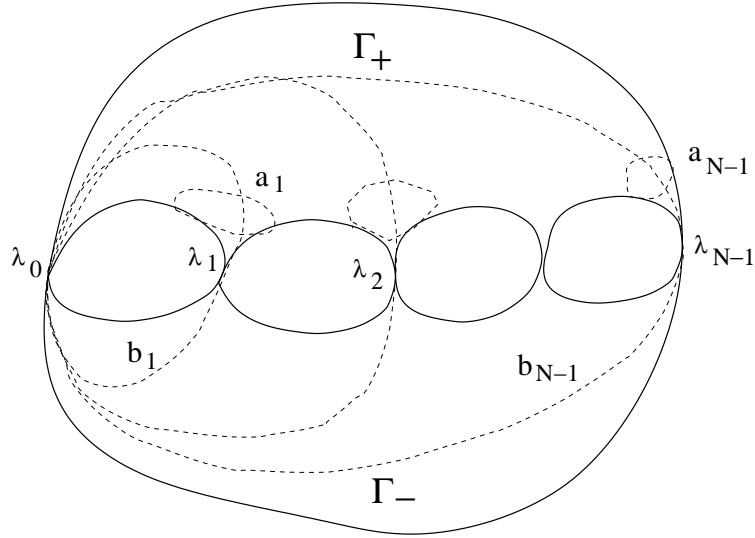}
\caption{Basis of circles}
\end{figure}
We define differentials $\omega_k$ by the formula
$$
\omega_k=\[ \frac{1}{ z-\l_k}- \frac{1}{ z-\l_0}\]d z,\qquad\qquad k=1,\ldots,N-1.
$$
Evidently, the normalization condition holds
$$
\int_{a_p}\omega_k=2\pi i \; \delta_{k}^p,\qquad\qquad \qquad\qquad k,p=1,\ldots, N-1;
$$
while $b$--periods of $\omega$'s are real and  infinite.
For the variables $\t$, we have
\bea
\t_k& = & \pi i k +  \log q(\l_k) - \log q(\l_0)\nonumber \\
&= & \pi i k + \sum\limits_{s=1}^{N-1} \int\limits^{\g_s}_{\infty_-}
\[\frac{1}{ z-\l_k}- \frac{1}{ z-\l_0}\]d z\\
&= & \pi i k + \sum\limits_{s=1}^{N-1} \int\limits^{\g_s}_{\infty_-} \omega_k.\nonumber
\eea
Therefore,  $\t$'s are the values of the Abel map
from the divisor $\g_1,\ldots< \g_{N-1}$ into $R^{N-1}$,  the noncompact real part of the Jacobian.
It will be shown that this map is onto.

Fix some matrix $L_0$. All matrices with the same spectrum  $\l_0<\ldots < \l_{N-1}$ as $L_0$
constitute a spectral class of $L_0$,  which we denote by $\SL$.
The spectral class $\SL$ is
in $1\,:\,1$ correspondence with  Weyl functions from $\Rn'$. 
$$
w(\l)= \sum\limits_{n=0}^{N-1} \frac{\rho_n}{ \l_n-\l},\qquad\qquad \qquad \sum \rho_n=1.
$$
 
\begin{thm} (i) The variables $\g_1,\ldots,\g_{N-1}$ are coordinates on $\SL$. Any sequence of $\g$'s
which occupies open segments $\l_{k-1}<  \g_k < \l_k,\qquad k=1,\ldots, N-1$ corresponds to some
matrix from $\SL$.
 
(ii) The variables $\t_1,\ldots, \t_{N-1}$ are coordinates on $\SL$. Any sequence of $\t$'s from
$R^{N-1}$ corresponds to some matrix from $\SL$.
\end{thm}
 
{\it Proof.} (i) The variables $\g$'s determine the roots of $q(\l)$ and therefore the function
$w(\l)$. Whence, $\g$'s are coordinates.
 
To prove that $\g$'s are free pick any sequence of $\g$'s and form $q(\l)$.  
Then, by Lagrange interpolation
$$
-\frac{q(\l)}{ p(\l)} =\sum\limits_{n=0}^{N-1} \frac{q(\l_n)}{ p'(\l_n)} 
\frac{1}{ \l_n- \l}. 
$$
It is easy to check all $\rho_n= q(\l_n)/ p'(\l_n)$ are strictly positive. It remains to prove
that $\sum \rho_n=1$. Indeed, the formula 
$$
q(\l)= \sum\limits_{n=0}^{N-1} \frac{q(\l_n)}{ p'(\l_n)}\frac{p(\l) }{ \l-\l_n}
$$
implies
$$
1=\lim\limits_{\l \rightarrow \infty} \frac{q(\l)}{ \l^{N-1}} =\lim\limits_{\l \rightarrow \infty}
\sum \frac{q(\l_n)}{ p'(\l_n)} \frac{p(\l) }{ \l^{N-1}(\l-\l_n)}= \sum  
\frac{q(\l_n)}{ p'(\l_n)}
$$
We are done.
 
(ii) From the definition of $\t$'s  for any $k=1,\ldots, N-1$:
$$
e^{\t_k}= (-1)^k \frac{q(\l_k)}{ q(\l_0)}.
$$
Using $q(\l_k)=p'(\l_k) \rho_k$ we have
$$
\rho_k=\rho_0 e^{\t_k} (-1)^k \frac{p'(\l_0)}{ p'(\l_k)}.
$$
It is easy to check that all $\rho_k >0$. There exists just one $\rho_0$ 
such that
$\sum\rho_n =1$. This implies that $\t$'s are coordinates and they are free. 
The Theorem  is proved.

The fact that the isospectral set  $\SL$ is a diffeomorphic to $R^{N-1}$ 
was already noted by Moser, \cite{MO}. Tomei, \cite{To} then showed that 
$\SL$ can be compactified and it becomes a convex polyhedron. 
The symplectic interpretation 
of this result as a version of  the Atiyah--Guillemin--Sternberg
convexity theorem was given by Bloch, Flaschka  and Ratiu \cite{BF}.

Now for a function  $w(\l) \in Rat'_N$ with $\l_0=0$ we consider another, 
exponential representation \ref{kss}: 
$$
w(\l)=-\frac{1}{ \l} \exp\,  \Xi(\l),
$$
and define another set of angles by the formula 
\beq\label{angn}
\t_k'= \lim_{\l\rightarrow \l_k}\[\Xi(\l)- \Xi(0) +\log\frac{\l_k-\l}{ \l_k}\] + \pi i
\qquad\qquad\qquad k=1,\ldots,N-1.
\eeq
This formula can be put in the form 
$$
\t_k'=\sum\limits_{s=1, s\neq k}^{N-1} \int\limits^{\gamma_s}_{\l_s}\omega_k +
\int\limits_{\infty_-}^{\gamma_k} \omega_k +\pi i.
$$
Whence, $\t_k'$ correspond to the Abel sum in the Baker form. The regularization is necessary, because 
the term $\int_{\l_k}^{\g_k} \omega_k$ diverges logarithmicaly on singular curve $\Gamma$. 
Fathermore,  we have simple relation 
\beq\label{rel}
\t_k=\t_k' + \pi i (k-1)   + \log \prod\limits_{s=1,  s\neq k}^{N-1} 
\frac{\l_s-\l_k}{ \l_s}.
\eeq
The angles $\t$ and $\t'$ differ by the real quantity which depend on the curve {\it only}. 
Evidently the variables $\t'$ are coordinates on $S(L_0)$ and their range is $R^{N-1}$.

\section{The Poisson Bracket on Weyl Functions.}
 
Following \cite{KK} we consider  functions $w(\l)$ with the properties {\it i.}
analytic in the
half--planes $\Im z > 0$ and $\Im z < 0 $. {\it ii.}  $w(\bar{z}) = \overline{w(z)}$, if
$\Im z \neq 0$
{\it iii.} $\Im w(z) > 0, $ if $ \Im z >0$. All such function are called $R$--functions. They play
central role in the theory of resolvent of selfadjoint operators. 
The Weyl function of a Jacobi matrix is an $R$--function. 

Atiyah and Hitchin  \cite{AH} introduced a Poisson structure on the space of 
rational functions.  In the recent paper \cite{FG} Faybusovich and 
Gehtman wrote  Atiyah-Hitchin and higher Poisson structures on rational
functions in  compact invariant form.  They defined the Atiyah-Hitchin
bracket  by the formula
\beq\label{ps}
\{w(\l),w(\m)\}=\frac{(w(\l)-w(\m))^2 }{ \l- \m }.
\eeq
Here we discuss  some of its' remarkable properties, \cite{V}. 

We think about $w(\l)$ as an element of some commutative complex algebra 
which depends holomorphicaly on the parameter $\l$. 
Evidently, \ref{ps} is skew--symmetric with respect to $\l$ and $\m$. 
It is natural require linearity of the bracket   
\beq\label{lin}
\{aw(\l)+b w(\l),w(\nu)\}=a\{w(\l),w(\nu)\} +b\{w(\l),w(\nu)\},
\eeq
where $a$ and $b$ are constants. 
The symbol $w(\l)$ for $\l$ inside the contour $C$ is 
given by the Cauchy formula
$$
w(\l)=\frac{1}{ 2\pi i} \int\limits_{C} \frac{w(\zeta)}{\zeta-\l} d \zeta.
$$
Whence due to \ref{lin} the values of the bracket in different points are 
related
$$
\{w(\l),w(\m)\}=\frac{1}{ 2\pi i} \int\limits_{C} \frac{\{w(\zeta),w(\m)\}}{
\zeta-\l} d\zeta.
$$
It can be verified that the bracket \ref{ps} satisfies this compatibility 
condition.

Also it is naturally to require for the bracket the Leibnitz rule
\beq\label{leib}
\{w(\l)w(\m),w(\nu)\}=w(\l)\{w(\m),w(\nu)\} + w(\m)\{w(\l),w(\nu)\}.
\eeq
It can be verified in a long but simple calculation that \ref{lin}--\ref{leib}
imply the Jacobi identity
$$
\{w(\l),\{w(\mu),w(\nu)\}\}+\{w(\m),\{w(\nu),w(\l)\}\}+
\{w(\nu),\{w(\l),w(\mu)\}\}=0.
$$

The particular useful to us is an invariance of \ref{ps}  under the group of 
linear--fractional transformations
\beq\label{lintr}
w\qquad \longrightarrow \qquad w'=\frac{aw+ b}{ cw + d},
\eeq
where $a,b,c,d$ are constants.
This property will be used in the construction of the second system of 
canonical  coordinates.

In our study of finite Jacobi matrices we need a small subclass 
$\Rn \subset R$. These are the functions given by the formula  \ref{repr}. 
All such   functions  have asymptotic expansion at infinity
$$
w(\l)= -\frac{s_0}{ \l} -\frac{s_1}{ \l^2} -\ldots.
$$
We consider submanifold $\Rn'$ with $s_0=1$, or    equivalently,    
$\sum \rho_k=1$. 
We will demonstrate that the Dirac restriction of the bracket \ref{ps} to this 
submanifold takes the form  
\beq\label{rps}
\{w(\l), w(\m)\}'=
(w(\l)-w(\m)) \(\frac{w(\l)-w(\mu)}{ \l -\mu} - w(\l) w(\mu)\).
\eeq

The linear 
Poisson structure on the phase space $\mathcal L$ is defined by the formulae
\beq\label{lp}
\{c_k,v_k\}=-c_k/2, \qquad\qquad\qquad \{c_k,v_{k+1}\}=c_k/2,
\eeq
and all other brackets vanish. 
Whence, the linear bracket \ref{lp}   corresponds to the 
restriction  \ref{rps} of the AH structure on  the submanifold $\Rn'$.

The formula \ref{rps} can be used to define the Poisson structure \ref{lp}. 
For example,
substituting \ref{re} into \ref{rps},  after simple algebra we obtain
$$
\frac{2c_0\{v_0,c_0\}}{ \l^2\m^2}\times\( \frac{1}{ \m} - \frac{1}{\l}\) + \ldots =
\frac{c_0^2}{ \l^2 \m^2} \times\(  \frac{1}{ \m} - \frac{1}{ \l}\) + \ldots .
$$
From this one can read  the first identity:  $\{c_0,v_0\}=-c_0/2$.

A construction of canonical coordinates for the bracket \ref{ps} or \ref{rps}  
will be given in terms of
various representations for  $\Rn$ and $\Rn'$.

\section{Canonical coordinates on $\Rn$. The Dirac reduction.}
We start with the construction of the first system of canonical coordinate on 
$\Rn$ for the bracket \ref{ps}.
The next Theorem shows that the  parameters 
$$
\l_0,\ldots,\l_{N-1};  \; \rho_0, \ldots, \rho_{N-1}
$$
in the formula  \ref{repr} are  "almost"  canonically paired. 
\begin{thm} The bracket  \ref{ps} in $\l-\rho$ coordinates has the form
\begin{eqnarray}
\{\rho_k,\rho_n\}&=&\frac{2\rho_k \,\rho_n}{ \l_n-\l_k}(1-\delta_k^n),\label{rr}\\
\{\rho_k,\l_n\}&=& \rho_k \delta_k^n, \label{rl}\\
\{\l_k,\l_n\}&=& 0 \label{ll}.
\end{eqnarray}
\end{thm}

{\it Proof.} We represent $\rho$'s and $\l$'s as contour integrals
$$
\rho_k=- \frac{1}{ 2\pi i} \int_{O_k} w(\zeta)\,d\zeta
$$
$$
\rho_k\; \l_k= - \frac{1}{ 2\pi i} \int_{O_k} \zeta w(\zeta)\,d\zeta 
$$
In both formulae the contour $O_k$ surrounding $\l_k$ is traversed counterclockwise. Therefore, for 
$k\neq n$

\begin{eqnarray}
\{\rho_k,\rho_n\} &=& \{\frac{1}{ 2\pi i} \int_{O_k} w(\zeta)\,d\zeta,
                        \frac{1}{ 2\pi i} \int_{O_n} w(\eta)\,d\eta\}=
\frac{1}{ (2\pi i)^2} \int_{O_k}\int_{O_n}  \{w(\zeta),w(\eta) \}\,d\zeta\,d\eta \nonumber\\
&=& \frac{1}{ (2\pi i)^2} \int_{O_k}\int_{O_n}\frac{(w(\zeta)-w(\eta))^2}{ \zeta- \eta}d\zeta\;
d\eta\nonumber\\
&=& \frac{1}{ 2\pi i} \int_{O_k} d\zeta w^2(\zeta)\[\frac{1}{2\pi i}\int_{O_n} d\eta
\frac{1}{ \zeta-\eta}\]\nonumber \\
&-& \frac{2}{ (2\pi i)^2} \int_{O_k}\int_{O_n}\frac{w(\zeta)w(\eta)}{ \zeta- \eta}d\zeta\; d\eta
\label{cross}\\
&+&\frac{1}{ 2\pi i}\int_{O_n} d\eta w^2(\eta)\[\frac{1}{ 2\pi i}\int_{O_k} d\zeta \frac{1}{ \zeta-\eta}\].
\nonumber
\end{eqnarray}
The two terms with  square brackets vanish and for \ref{cross} we obtain
\beq\label{cro}
-\frac{2}{ 2\pi i}\int d\zeta w(\zeta) \[\frac{1}{ 2\pi i}\int d\eta\frac{w(\eta)
}{ \zeta-\eta}\].
\eeq
For $\zeta$ in the exterior of the contour $O_p$ we have
$$
\frac{1}{2\pi i}\int_{O_p} d\eta \frac{w(\eta)}{\zeta-\eta}=\frac{\rho_p}{\l_p-\zeta}.
$$
Applying this formula twice to \ref{cro} we obtain \ref{rr}. 
If $k=n$  then similar arguments show that the bracket \ref{rr} vanishes.

To prove \ref{rl} we compute for $k\neq n$
$$
\{\l_k\rho_k,\rho_n\}=\frac{2\l_k \rho_k \rho_n }{ \l_n-\l_k}.
$$
From another side, 
$$
\{\l_k\rho_k,\rho_n\}= \l_k\{ \rho_k,\rho_n\} + \rho_k \{\l_k,\rho_n\}.
$$
This together with \ref{rr} imply that the bracket $\{\l_k,\rho_n\}$ vanish.  
For $k=n$ we have  \newline
$\{\l_k\rho_k,\rho_k\}=-\rho^2_k$ and $\{\rho_k,\l_k\}=\rho_k$. 
The formula \ref{rl} is proved.

To prove \ref{ll} for $k\neq n$ we compute
$$
\{\l_k\rho_k,\l_n \rho_n\}=\frac{2\l_k\rho_k\l_n\rho_n}{ \l_n-\l_k}.
$$
From another side,  
$$
\{\l_k\rho_k,\l_n \rho_n\}=\rho_k \rho_n  \{\l_k,\l_n \}+ \rho_k \l_n \{\l_k, \rho_n\}
+\rho_n\l_k \{\rho_k,\l_n \} +\l_k \l_n \{\rho_k, \rho_n\}. 
$$
This together with formulae \ref{rr}-\ref{ll} imply that the bracket $\{\l_k,\l_n\}$ vanishes. 
For $k=n$ arguments are the same. The proof is finished.

Therefore,  for the bracket \ref{ps}   on $Rat_N$ canonical  coordinates  are 
associated with  poles of $w(\l)$: 
$$
\l_0,\ldots,\l_{N-1}, q(\l_0),\ldots, q(\l_{N-1}).
$$
Indeed, from the identity
$$
-\frac{q(\l)}{ p(\l)}=\sum\frac{\rho_n}{ \l_n-\l}
$$ 
and condition $p(\l_n)=0$ we obtain
$
q(\l_k)=p'(\l_k)\rho_k.
$
Furthermore, using \ref{rl}-\ref{ll} 
$$
\{q(\l_k),\l_n\}=\{p'(\l_k)\rho_k,\l_n\}=p'(\l_k)\rho_k\delta_k^n=q(\l_k)\delta_k^n.
$$
All other brackets vanish: 
$$
\{q(\l_k),q(\l_n)\}=\{\l_k,\l_n\}=0.
$$
In this coordinate form the Poisson structure on $Rat_N$ was introduced in \cite{AH}. These  
identities imply
$$
\{q(\l),q(\m)\}=\{p(\l),p(\m)\}=0
$$
and
$$
\{q(\l),p(\m)\}=\frac{q(\l)p(\m)-q(\m)p(\l)}{\l-\m}.
$$
The last expression is called a Bezoutian, see \cite{KN}.  This form of the bracket easily leads to 
\ref{ps}. 

The second set of canonical coordinates on $\Rn$ is associated with zeros of 
$w(\l)$:
$$
\g_1,\ldots \g_{N-1}; \; p(\g_1),\ldots, p(\g_{N-1});\; q_0,p_0.
$$
To prove this, we introduce the new "dual" function $w'(\l)$ as 
an inverse of the function \ref{det}
$$
w'(\l)=-\frac{1}{ w(\l)}=\frac{p(\l)}{ q(\l)}.
$$
Due to \ref{lintr} 
the bracket for the dual function is given by the formula \ref{ps}.
The new meromorphic function maps  the upper half-plane into itself and has the
expansion
$$
w'(\l)= \frac{\l}{ q_0} + c+ \sum\limits_{s=1}^{N-1} \frac{\rho_s'}{ \g_s-\l},
$$
where,
$$
c=\frac{p_0q_0-q_1}{ q_0^2}=\frac{p_0}{ q_0}+ \frac{\sum\gamma_s}{ q_0},\qquad \rho'_s >0.
$$

\begin{thm}  The following identities hold
\begin{eqnarray}
\{\rho'_k,\rho'_n\}& =& \frac{2\rho'_k\rho_n'}{ \gamma_n-\gamma_k} (1-\delta^n_k),\label{rrr}\\
\{\rho_k',\gamma_n\}&=& \rho_k'\delta_k^n,\label{rg}\\
\{\gamma_k,\gamma_s\}&=&0,  \label{gg}\\
\{q_0,\rho_k'\}&=&\{q_0,\gamma_s\}=0,\label{ccc}\\
\{\rho_k',p_0\}&=& \rho_k'\label{wro}\\
\{p_0,\gamma_s\}& =& 0,\label{zzz}\\
\{p_0,q_0\}&=& q_0.  \label{pop}
\end{eqnarray}
\end{thm}

{\it Proof.} The identities \ref{rrr}--\ref{gg} using integral representation
 
$$
\rho_k'=- \frac{1}{ 2\pi i} \int_{O_k} w'(\zeta)\,d\zeta
$$
$$
\rho_k'\; \gamma_k= - \frac{1}{ 2\pi i} \int_{O_k} \zeta w'(\zeta)\,d\zeta
$$
can be proved exactly the same way as identities \ref{rr}--\ref{ll} of
Theorem 2.
 
Let us compute the first bracket \ref{ccc}
\begin{eqnarray}
\{q_0,\rho_k'\}&=&\{\lim_{\l\rightarrow \infty} \frac{\l}{ w'(\l)}, -\frac{1}{ 2\pi
i} \int_{O_k}
w'(\zeta) d\zeta\}\nonumber\\
               &=& \lim_{\l\rightarrow \infty} \frac{\l}{  2\pi i w'(\l)^2} \int_{O_k}
\frac{(w'(\l)-w'(\zeta))^2}{ \l - \zeta}\; d\zeta. \nonumber
\end{eqnarray}
Expanding the square and computing each term separately we see that the bracket
vanishes. The proof of the second identity  \ref{ccc} is exactly the same.
 
To prove \ref{wro} we note
$$
\{c,\rho'_k\}=\frac{\{p_0,\rho_k'\}}{ q_0}- \frac{\rho_k'}{ q_0}.
$$
From another hand
\begin{eqnarray}
\{c,\rho_k'\}&=&\{\lim_{\l\rightarrow \infty}  w'(\l)-\frac{\l}{ q_0}, -\frac{1}{ 2\pi i} \int_{O_k}
w'(\zeta) d\zeta\}\nonumber\\
               &=& \lim_{\l\rightarrow \infty} - \frac{1}{ 2\pi i} \int_{O_k}
\frac{(w'(\l)-w'(\zeta))^2}{ \l - \zeta}\; d\zeta \nonumber\\
&=& -\frac{2\rho_k'}{ q_0}.\nonumber
\end{eqnarray}
Comparing it with the previous formula we obtain \ref{wro}. The proof of the formula \ref{zzz}
is similar.
 
To prove the last formula \ref{pop} we compute
\begin{eqnarray}
\frac{\{p_0,q_0\}}{ q_0} &=& \{c,q_0\}=\{ \lim_{\l\rightarrow \infty} w'(\l)-
\frac{\l}{ q_0}, q_0 \}\nonumber\\
                      & =& \lim_{\l\rightarrow \infty} \{w'(\l),\lim_{\m\rightarrow \infty}
\frac{\mu}{ w'(\mu)} \} = \lim_{\l\rightarrow \infty}  \lim_{\m\rightarrow \infty} -
\frac{\mu}{ w'(\mu)^2} \frac{(w'(\l)-w'(\m))^2}{ \l-\m}= 1.
\nonumber
\end{eqnarray}
This implies the result. The proof is finished.

From \ref{rg}  using the formula $p(\g_s)=-\rho_s' q'(\gamma_s)$ we obtain
$$
\{p(\g_n),\g_k\}= p(\g_n)\delta_n^k.
$$
From \ref{rrr}, \ref{rg} and \ref{ccc}
$$
\{p(\g_n),p(\g_k)\}=0.
$$
From \ref{ccc}
$$
\{p(\g_n), q_0\}=0.
$$
From \ref{wro} and \ref{pop}
$$
\{p(\g_n),p_0\}=0.
$$
These identities together with identities of  Theorem  3  provide a proof  of
our  statement.

This coordinate system is usefull in construction of the Dirac restriction, 
\cite{Dir},  of the AH bracket \ref{ps} on the submanifold $M\subset \Rn$ 
determined by the conditions  
$$\Phi_1=p_0=c_1,\qquad\qquad\qquad\Phi_2=\log q_0=c_2$$
where $c_1$ and $c_2$ are some real constants.  

Consider a more general problem. The submanifold $M$ of 
dimension $2N-m$ is determined by the conditions 
$$\Phi_1=c_1,\;\;\Phi_2=c_2, \qquad\hdots \qquad\qquad, \Phi_m=c_m,$$
where $\Phi$'s are some functions on the phase space and $c$'s are 
real constants. The bracket $\{\bullet,\bullet\}$ is  modified
$$
\{F_1,F_2\}'=\{F_1,F_2\}+ \sum\limits_{k=1}^{m} \sigma_k \{F_1,\Phi_k\}
$$
whith  $\sigma$'s chosen  such that 
$$
\Phi_k^{\bullet}=\{\Phi_k,F_2\}+\sum\limits_{s=1}^{m} \sigma_s 
\{\Phi_k,\Phi_s\}=0. 
$$
for all $k=1,\hdots, m$. Geometricaly, this condition means that the vector 
fields produced in the bracket $\{\bullet,\bullet\}'$ are tangent to $M$. 
If the matrix $\|\{\Phi_k,\Phi_s\}\|$ has an inverse $\|C_{ks}\|$, then the 
last system can be solved for $\sigma$'s and 
$$
\{F_1,F_2\}'=\{F_1,F_2\}+ \sum\limits_{k, s=1}^{m}  \{F_1,\Phi_k\} \, C_{ks}\,  
\{F_2,\Phi_s\}. 
$$

Implementing this procedure for our choice of functionals $\Phi_1$ and $\Phi_2$ 
we obtain 
$$
\sigma_1=\{\log q_0,F_2\},\qquad\qquad \sigma_2= -\{p_0,F_2\}
$$
and
$$
\{F_1,F_2\}'=\{F_1,F_2\} +     
\{\log q_0, F_2\}  \{F_1,p_0\}-\{p_0, F_2\} \{F_1,\log q_0 \}.     
$$
One can easily verify 
$$
\{q(\l),q(\m)\}'= \{p(\l),p(\m)\}'=0.
$$
Using $\{p_0,q(\l)\}=q(\l)$ and $\{p(\m),q_0\}=q(\m)$ we obtain 
$$
\{q(\l),p(\m)\}'=\frac{q(\l)p(\m)-q(\m)p(\l)}{ \l-\m} +\frac{q(\m)q(\l)}{ q_0}. 
$$
Finaly, we have
$$
\{w(\l), w(\m)\}'=
(w(\l)-w(\m)) \(\frac{w(\l)-w(\mu)}{ \l -\mu} - \frac{w(\l) w(\mu)}{ q_0} \).
$$
This becomes \ref{rps} for a particular choice $\Phi_2=\log q_0=0$.

\section{The  canonical coordinates on $\Rn'$.}

Now we turn to the submanifold $Rat_N'$ with  $q_0= \sum\rho_k=1$ 
and the Poisson bracket \ref{rps}. 
Here the situation is a little more subtle.
In all formulae we omit the prime near the bracket $\{\bullet,\bullet\}'$.  

\begin{thm} The bracket  \ref{rps} in $\l-\rho$ coordinates has the form
\begin{eqnarray}
\{\rho_k,\rho_n\}&=&\[\frac{2\rho_k \,\rho_n}{ \l_n-\l_k} - 2 \rho_k\rho_n\(\sum\limits_{s\neq k} 
\frac{\rho_s}{\l_s-\l_k} - \sum\limits_{s\neq n} 
\frac{\rho_s}{ \l_s-\l_n}\)\] 
(1-\delta_k^n),\label{drr}\\
\{\rho_k,\l_n\}&=& -\rho_k\rho_n + \rho_k \delta_k^n, \label{drl}\\
\{\l_k,\l_n\}&=& 0 \label{dll}.
\end{eqnarray}
\end{thm}
Proof is similar to the proof of Theorem 2 and therefore is omitted. 

The Theorem implies,
\beq\label{bcr}
\{q(\l_k),\l_n\}=-q(\l_k)\rho_n+q(\l_k)\delta_k^n.
\eeq
Thus, we have the first system of canonical coordinates, so--called, 
the action-angle variables  
\beq\label{naa}
\l_1,\ldots,\l_{N-1};\t_1,\ldots,\t_{N-1} 
\eeq
where $\t$'s  defined by \ref{angl} are real and canonically paired  with 
$\l$'s.  Indeed,  \ref{bcr} implies
$$
\{\t_k ,\l_n\}=(\delta_k^n-\delta_0^n).
$$
The formulae \ref{drr} and \ref{bcr} produce
\beq\label{qcro}
\{\log q(\l_k), \log q(\l_n)\}= \sum\limits_{s\neq k} \frac{\rho_s +\rho_k}{ \l_k-\l_s} -
                               \sum\limits_{s\neq n} \frac{\rho_s +\rho_n}{ \l_n-\l_s}.
\eeq
This  identity implies the commutativity of the angles:  $\{\t_k,\t_n\}=0$.

Using  the Theorem  it can be checked easily that $p_0=-\sum\l_k$ is a
Casimir of the bracket 
\ref{rps}.


Evidently, for the restricted bracket  \ref{rps}  the canonical relations
established in Theorem 3 survive.
We have the second set   of canonical variables  on $\Rn'$
\beq\label{ndq}
\g_1,\ldots,\g_{N-1},\p_1,\ldots,\p_{N-1};\qquad\qquad\qquad\qquad\p_k=\log (-1)^{N+k} p(\g_k).
\eeq
These divisor-quasimomentum coordinates were introduced in \cite{KV}. 
The denominator $p(\l)=(-1)^N\prod(\l_n-\l)$ satisfies
$$
(-1)^{N+k}p(\g_k)>0,\qquad\qquad k=1,\ldots,N-1.
$$
Whence $\p$'s  are real and  canonicaly paired with $\g's$:
$$
\{\p_n,\g_k\}=\d_n^k.
$$
All other brackets vanish.

In the rest of this section we show that the variables \ref{angn} 
associated with   representation of 
$w(\l) \in Rat_N'$  in the exponential form \ref{kss} 
$$
w(\l)=-\frac{1}{ \l} e^{\Xi(\l)} 
$$
can be moved by corresponding  $\l$'s:
$$
\{\theta_k',\l_n\}=\delta_n^k,\qquad\qquad\qquad\qquad k,n=1,\ldots,N-1.
$$
Though it   follows  from the previous discussion of the 
action-angle variables  and the 
formula \ref{rel} we will give an independent proof of this fact.  It is 
important to notice that we can not expect commutativity of the variables 
$\theta_n'$.

The multivalued function $\Xi(\l)$ has the form
$
\Xi(\l)=\sum_{s=1}^{N-1} \log(\g_s-\l) -\log(\l_s-\l)
$ and defined up to an integer multiple of $2\pi i$. The bracket $\{\;\bullet\;, \Xi(\l)\}$ is 
single valued since additive constant vanishes.  
The Poisson bracket \ref{rps} in terms of the function $\Xi(\l)$ has the form
$$
\{\Xi(\l),\Xi(\m)\}=\frac{4\sinh^2\(\Xi(\l)-\Xi(\m)-\log\l +\log \m/2\)}{ \l-\mu} + 
\frac{1}{ \l}e^{\Xi(\l)}-\frac{1}{ \m}e^{\Xi(\m)}
$$
or 
\beq\label{rpsx}
\{\Xi(\l),\Xi(\m)\}=\frac{1}{ w(\l) w(\m)}  \frac{\(w(\l)-w(\mu)\)^2}{ \l -\mu} - w(\l)+ w(\mu)
\eeq
which is more convinient for calculations.  The pole $\l_k$  can be represented as a contour 
integral
$$
\l_k= -\frac{1}{ 2\pi i} \int\limits_{O_k} \zeta\;  d\,   \Xi(\zeta) =-\l_k' + 
\frac{1}{ 2\pi i} \int\limits_{O_k} \,  \Xi(\zeta) \,d\zeta, 
$$
where $\l_k'$ is an arbitrary fixed point on the contour $O_k$ surrounding $\l_k$. 
As a simple example we   prove commutativity of $\l$'s 
$$
\{\l_k,\l_n\}= \frac{1}{ (2\pi i)^2} \int\limits_{O_k}\int\limits_{O_n}  
\{\Xi(\zeta),\Xi(\eta) \}\,d\zeta\,d\eta 
$$
From  the formula \ref{rpsx} one can easily see that 
the double integral vanishes.  

Now for the angles $\t'_n$ we have 
$$
\{\t_k',\l_n\}=\lim_{\l\rightarrow \l_k} \[\{\Xi(\l),\l_n\} -\{\Xi(0),\l_n\} + 
\{\log \l_k-\l/ \l_k,\l_n\}\]
$$
The last term vanishes. The first two terms are more complicated
\begin{eqnarray}
\{\Xi(\l),\l_n\} &=& 
\frac{1}{ 2\pi i} \int\limits_{O_n}  \{\Xi(\l),\Xi(\zeta) \}\,d\zeta\ \nonumber\\
&=& \frac{1}{ 2\pi i} \int\limits_{O_n}\frac{w(\l)}{ w(\zeta)} 
\frac{d\zeta}{ \l- \zeta}\;
\label{one}\\
&-& \frac{2}{ 2\pi i} \int\limits_{O_n} \frac{d\zeta}{ \l -\zeta}\label{two} \\
&+& \frac{1}{ 2\pi i} \int\limits_{O_n}\frac{w(\zeta)}{ w(\l)} \frac{d\zeta}{ \l- \zeta} 
\label{three}\\
&-&\frac{1}{ 2\pi i}\int\limits_{O_n}[w(\l)-w(\zeta)]d\zeta.
\label{four}
\end{eqnarray}
If   $\l\rightarrow \l_k,\quad k\neq n, $ or   $\l\rightarrow \l_0=0$,      
the terms \ref{one}-\ref{three} vanish. 
The term \ref{four} is equal to $-\rho_n$.  Therefore,
$$
\lim_{\l\rightarrow \l_k} \{\Xi(\l),\l_n\}=-\rho_n,\quad\qquad k\neq n;\qquad
\{\Xi(0),\l_n\}=-\rho_n.
$$
This implies $\{\t_k',\l_n\}=0,\quad k\neq n$. 

Furthermore, if $\l\rightarrow \l_n$, then the term 
\ref{one} becomes $-1$, the term \ref{two} becomes 2 and \ref{three} vanishes. The term \ref{four} 
is $-\rho_n$. Therefore,
$$
\lim_{\l\rightarrow \l_n} \{\Xi(\l),\l_n\}=1 -\rho_n
$$
Thus $\{\t_n',\l_n\}=1$.

\section{Tangent and Transversal Flows.}
Using the poles of $w(\l)$ we define Hamiltonians $H_j=\frac{1}{ j} \sum \l_n^j,\quad j=1,\ldots,N$. 
The flows produced by them in the bracket \ref{rps} are tangent to the isospectral manifold: 
$\{\l_k, H_j\}=0.$  Due to Theorem  4 the  standard, \cite{MO}, 
Toda flows have the form 
$$
\rho_k^\bullet=\{\rho_k,H_j\}=\(\l_k^{j-1}-\sum\l_n^{j-1} \rho_n\) \rho_k,\qquad\qquad\qquad 
k=0,\ldots,N-1.
$$ 
Toda flows commute $\{H_j,H_k\}=0$ and linearized in the variables \ref{naa}: 
$$
\t_k^{\bullet}=\{\theta_k, H_j\}=(\l_k^{j-1}-\l_0^{j-1}),\qquad\qquad\qquad\qquad k=1,\ldots,N-1. 
$$

Similarly, from zeros of $w(\l)$ we define another set of Hamiltonias $T_j=
\frac{1}{ j}\sum\g_n^j, 
\quad j=1,\ldots,N-1$. By  Theorem  3  the flows produced by these 
Hamiltonians do not affect $\g$'s. Therefore, we call these  commuting 
flows transversal. They are linearized in the variables 
\ref{ndq}
$$
\p_k^\bullet=\{\p_k,T_j\}=\g_k^{j-1},\qquad\qquad\qquad\qquad   k=1,\ldots,N-1.
$$

This is an example of the situation similar to the one considered in 
physics, \cite{FGN, BMM}. Given two systems of canonical coordinates 
and two families 
of commuting Hamiltonians. Each family depends only on the half of coordinates 
of the corresponding canonical system. Hamiltonians of both families produce coordinate system for the Poisson manifold. 

It is routine exercise to derive the equations of motion for $\g$'s under 
$H$ flows and for $\l$'s under $T$ flows. Then the inverse spectral problem can be solved using trace formulae of section 2. We do not dwell on this.

\end{document}